\documentclass{article}
\usepackage{amsmath,amssymb}
\usepackage{graphicx}
\usepackage{tikz}

\def\EE{\mathbb E}

\def\RR{\mathbb R}

\def\ZZ{\mathbb Z}

\def\cB{\mathcal B}

\def\cL{\mathcal L}
\def\cS{\mathcal S}
\def\cP{\mathcal P}

\def\bx{\mathbf x}
\def\be{\mathbf e}

\def\bx{\mathbf x}
\def\by{\mathbf y}
\def\ba{\mathbf a}

\def\bc{\mathbf c}
\def\bb{\mathbf b}
\def\bp{\mathbf p}

\def\bz{\mathbf z}

\def\b1{\mathbf 1}
\def\gl{\lambda}
\def\ga{\alpha}
\def\gb{\beta}

\def \mw{\mathfrak{w}}
\def \mb{\mathfrak{b}}
\def \mS{\mathfrak{S}}
\def \mP{\mathfrak{P}}

\newcommand{\qed}{\hfill$\square$\bigskip}
\newcommand{\raf}[1]{(\ref{#1})}
\newcommand{\proof}{\noindent {\bf Proof}~~}

\newcommand{\hide}[1]{}

\newcommand{\Max}{\textsc{Max}}
\newcommand{\Min}{\textsc{Min}}

\newtheorem{theorem}{Theorem}
\newtheorem{lemma}{Lemma}
\newtheorem{corollary}{Corollary}

\newtheorem{fact}{Fact}

\newtheorem{remark}{Remark}

\newtheorem{claim}{Claim}

\title{A Nested Family of $k$-total Effective Rewards for Positional Games}

\author{
Endre Boros\thanks{MSIS Department and RUTCOR, Rutgers University, 100 Rockafellar Road, Livingston Campus
Piscataway, NJ 08854, USA;
($\{$boros,gurvich$\}$@rutcor.rutgers.edu)}
\and
Khaled Elbassioni\thanks{Masdar Institute of Science and Technology, P.O.Box 54224, Abu Dhabi, UAE;
(kelbassioni@masdar.ac.ae)}
\and
Vladimir Gurvich\footnotemark[2]
\and
Kazuhisa Makino\thanks{Research Institute for Mathematical Sciences (RIMS)
Kyoto University, Kyoto 606-8502, Japan;
(makino@kurims.kyoto-u.ac.jp)}
}

\begin{document}

\maketitle

\date

\abstract{
We consider Gillette's two-person zero-sum stochastic games with perfect information.
For each $k \in \ZZ_+$ we introduce an effective reward function, called $k$-total.
For $k = 0$ and $1$ this function is known as {\it mean payoff} and {\it total reward}, respectively.
We restrict our attention to the deterministic case.
For all  $k$, we prove the existence of a saddle point which can be realized by uniformly optimal pure stationary strategies.  We also demonstrate that $k$-total reward games can be embedded into
$(k+1)$-total reward games.

{\bf Keywords}: stochastic game with perfect information, cyclic games, two-person, zero-sum, mean payoff, total reward}

\section{Introduction}

We consider two-person zero-sum stochastic games with perfect information
and for each positive integer $k$ define an effective reward function, called the $k$-total reward,
generalizing the classical \emph{mean payoffs} \cite{Gilette57} ($k = 0$), as well as the \emph{total rewards}
\cite{TV87,TV98} ($k = 1)$.

In this paper, we restrict ourselves to two-person zero-sum
games with deterministic positions,
and the solution concept is Nash equilibrium, which is just a saddle point in the considered case.
We call the considered family of games
\emph{$k$-total reward BW-games}, where
B and W stand for the two players: Black (the minimizer) and White (the maximizer);
in the sequel we shall denote them by \Min\ and \Max.

We denote by $\RR$ the set of reals, by $\ZZ$ the set of integers, and by $\ZZ_+$ the set of nonnegative integers.
For a subset $I\subseteq \ZZ_+$, let  $\RR^I$ denote the set of vectors indexed by the elements of $I$.
In particular, $\cS=\RR^{\ZZ_+\setminus\{0\}}$ denotes the set
of infinite real sequences. For $\ba \in\cS$ we write $\ba=(a_1,a_2,\ldots)$.
Furthermore, for $m\in \ZZ_+\setminus\{0\}$ we define $[m]=\{1,2,\ldots,m\}$ and write simply $\RR^m$ instead of $\RR^{[m]}$.

To describe BW-games, let us consider a finite directed graph (digraph) $G=(V,E)$, whose vertices
(also called positions or states) are partitioned into two sets $V=B\cup W$, a fixed initial
position $v_0\in V$, a function $r:E \rightarrow \RR$ assigning real numbers to the arcs (moves\footnote{Following standard terminology, we will use {\it vertices} and {\it arcs} when we talk about graphs and {\it positions} and {\it moves} when we talk about games.}), and a mapping $\pi:\cS\rightarrow \overline\RR$, where $\overline\RR=\RR\cup\{-\infty,+\infty\}$.
We call the tuple $(G,r,\pi)$ a \emph{BW-game} where $r$ is its \emph{local reward} and
$\pi$ is its \emph{effective reward}.
Two players, \Min\ and \Max\ control the positions of $B$ and $W$, respectively.
The game begins at time $t=0$ in the initial position $s_0=v_0$. In a general step, in time $t$, we are at position $s_t\in V$.
The player who controls $s_t$ chooses an outgoing arc $e_{t+1}=(s_t,v)\in E$, and the game moves to position $s_{t+1}=v$.
We assume, in fact without any loss of generality, that every vertex in $G$ has an outgoing arc.
(Indeed, if not, one can add loops to terminal vertices.)
We assume that an initial vertex $v_0$ is fixed. However, when we talk about solving a BW-game, we
consider  (separately) all possible initial vertices. In general, the strategy of the player is a policy by which (s)he chooses the outgoing arcs from the vertices (s)he controls. This policy may involve the knowledge of the previous steps as well as probabilistic decisions. We call a strategy {\it stationary} if it does not depend on the history and {\it pure} if it does not involve probabilistic decisions; for more details see Section~\ref{sec:3}.

In the course of this game players generate an infinite sequence
of edges $\bp=(e_1,e_2,\ldots)$ (a \emph{play}) and the corresponding real sequence
$r(\bp)=(r(e_1),r(e_2),\ldots)\in\cS$ of local rewards.
At the end (after infinitely many steps) \Min\ pays \Max\ $\pi(r(\bp))$ amount.
Naturally, \Max's aim is to create a play which maximizes $\pi(r(\bp))$, while \Min\ tries to minimize it.
(Let us note that
the local reward function $r:E\rightarrow \RR$ may have negative values, and
$\pi(r(\bp))$ may also be negative, in which case
\Max\ has to pay \Min.) 
As usual, a pair of (not necessarily pure or stationary) strategies is a saddle point if neither of the players can improve individually by changing her/his strategy. The corresponding $\pi(r(\bp))$ is the value of the game with respect to initial position $v_0$. Such a pair of strategies are called {\it optimal}; furthermore, it is called {\it uniformly optimal} if it provides the value of the game for any initial position.

As we shall see later, it will be enough to restrict ourselves, and the players, to their pure stationary strategies in these BW-games.
This means that each player chooses, in advance, a move in every position that
(s)he controls and makes this move whenever the play comes to this position.
Then, the play is uniquely determined by the one time selection of arcs and by the initial position.
Such a play always looks like a ``lasso": it consists of
an initial path entering a directed cycle, which is then repeated infinitely many times.

\bigskip

\emph{Mean payoff} (undiscounted) stochastic games, introduced in \cite{Gilette57}, include BW-games
(see also \cite{Mou76a,Mou76, EM79, GKK88}) with effective reward function $\pi=\phi$:
\begin{equation}\label{mean payoff}
\phi(\ba) ~=~ \liminf_{T\rightarrow \infty} \frac{1}{T} \sum_{j=1}^T a_j,
\end{equation}
where $\ba:=(a_1,a_2,\ldots)$ is the sequence of expected local rewards incurred at times $1,2,\ldots$ of the play.
Such a game is known to have a saddle point in pure stationary strategies
\cite{Gilette57,LL69}. 

\emph{Discounted (mean) payoff} stochastic games were in fact introduced earlier in \cite{Shapley53} and have payoff function $\pi=\phi_{\gb}$:
\begin{equation}\label{disounted mean payoff}
\phi_{\gb} (\ba) ~=~ (1-\gb) \sum_{j=1}^{\infty} \gb^{j-1} a_j,
\end{equation}
where $\gb\in[0,1)$ is the so-called {\it discount factor}. Since the set of positions $V$ is finite, the $a_j$ values are bounded. Thus, as a consequence of the classical Hardy-Littlewood Tauberian theorems \cite{HL31}, we have the equality
\begin{equation}\label{e-HL}
\phi(\ba) ~=~ \lim_{\gb \rightarrow 1^-} \phi_{\gb} (\ba).
\end{equation}
Discounted games, in general, are easier to solve, due to the fact that a standard value iteration is a converging contraction.
For this reason they are widely used in the literature of stochastic games together with the above limit equality.
In fact, for mean payoff BW-games with integral rewards with maximum absolute value $R$, it is known \cite{ZP96} that, if $1-\gb \leq \frac{1}{4|V|^3R}$ then for any two infinite reward sequences
$\ba,\bb$ the inequality $\phi(\ba) < \phi(\bb)$ implies $\phi_{\gb}(\ba)<\phi_{\gb}(\bb)$.

\bigskip

\emph{Total reward}, introduced in \cite{TV87} and considered in more detail in \cite{TV98}, is defined by
\begin{equation}\label{total reward}
\psi (\ba) ~=~ \liminf_{T\rightarrow \infty} \frac{1}{T} \sum_{i=1}^T \sum_{j=1}^i a_j.
\end{equation}
It was shown in \cite{TV98} that a total reward game is equivalent with a mean payoff game having countably many positions.
From this the authors derive that every total reward game has a value.
Furthermore, $\epsilon$-optimal Markovian strategies\footnote{A history-dependent strategy is called Markovian if the move depends only on time and position (but not the complete history).} can be constructed.
The proof of the latter is analogous to the proof in \cite{MN81}.

It is worth noting that the $1$-total reward games with nonnegative local rewards are polynomially solvable \cite{MSS04}.
This contrasts the fact that mean payoff games with nonnegative rewards are as hard as general mean payoff games, and that
the fastest known algorithms for mean payoff BW-games are either pseudo-polynomial \cite{GKK88, Pis99, ZP96} or
randomized subexponential \cite{BV05, Halman07, Vorobyov08}.

\medskip

In this paper we extend and generalize the above results.
For every $k \in \ZZ_+$, we define the $k$-total effective
reward, which coincides with the mean payoff when  $k = 0$  and
with the total reward when  $k = 1$. 

In general, given a sequence of local rewards $\ba=(a_1,a_2,\ldots)$ let us associate to it another sequence $M(\ba)=(a_1,a_1+a_2,\ldots)$. Then the $k$-total  
reward $\phi^{(k)}(\ba)$ is defined as the mean payoff of the sequence $M^k(\ba)$. Let us note that $M(\ba)$ may not be a bounded sequence, even if $\ba$ is. Consequently, $\phi^{(k)}$ may take infinite values. 

\paragraph{Examples.}
For instance, consider the following 5 sequences:
\begin{align*}
\ba^0&=(0,0,\ldots),\\
\ba^1&=(1,0,-1,0,1,0,-1,\ldots),\\
\ba^2&=(-1,0,1,0,-1,0,1,\ldots),\\
\ba^3&=(1,-1,-1,1,1,-1,-1,1,1,\ldots),\\
\ba^4&=(-1,1,1,-1,-1,1,1,-1,-1,\ldots).
\end{align*}
Then we have $M(\ba^4)=\ba^2$ and $M(\ba^5)=\ba^3.$
\begin{equation*}
\begin{array}{r|c|c|c|c|c}
&\ba^0&\ba^1&\ba^2&\ba^3&\ba^4\\
\hline
\phi^{(0)}(\ba^i)& 0 & 0 &0 &0 &0\\
\phi^{(1)}(\ba^i)& 0 & \frac{1}{2} &-\frac{1}{2} &0 &0\\
\phi^{(2)}(\ba^i)& 0 & +\infty &-\infty &\frac{1}{2} &-\frac{1}{2}\\
\phi^{(3)}(\ba^i)& 0 & +\infty &-\infty & +\infty &-\infty\\
\phi^{(4)}(\ba^i)& 0 & +\infty &-\infty & +\infty &-\infty
\end{array}
\end{equation*}

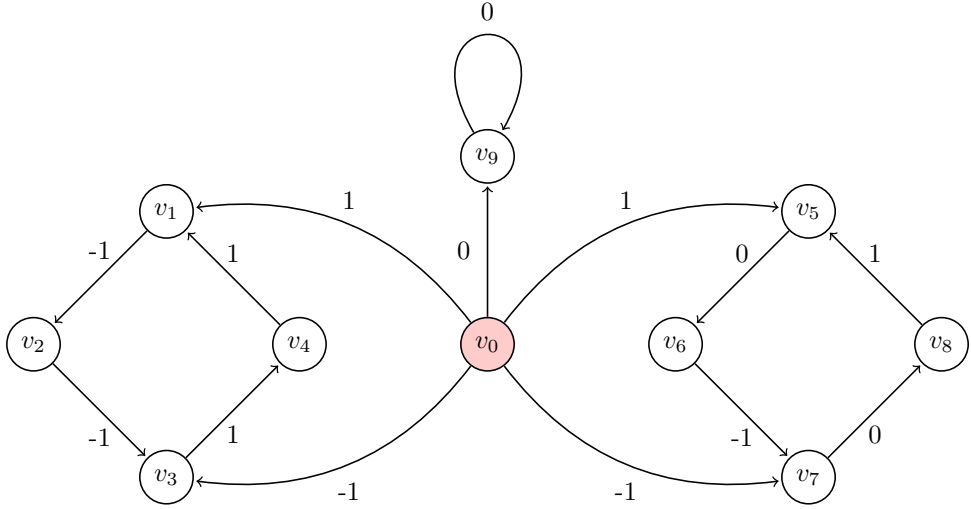
\begin{figure}
\begin{tikzpicture}[->,shorten >=1pt,auto,node distance=2.5cm,semithick]
\tikzstyle{every node}=[draw,shape=circle]
  
  \node[fill=red!20] (v0)  {$v_0$};
  \node (v9) [above of =v0] {$v_9$};
  \node (v4) [left of =v0] {$v_4$};
  \node (v1) [above left of =v4] {$v_1$};
  \node (v2) [below left of =v1] {$v_2$};
  \node (v3) [below left of =v4] {$v_3$};
  \node (v6) [right of =v0] {$v_6$};
  \node (v5) [above right of =v6] {$v_5$};
  \node (v7) [below right of =v6] {$v_7$};
  \node (v8) [below right of =v5] {$v_8$};
  
  \path[->,min distance=2cm]
  (v9) edge [above,in=60,out=120,looseness=10] node[draw=none] {0} (v9);
  \path
  (v0) edge node[draw=none] {0} (v9)
  (v0) edge[bend right] node[draw=none,above] {1} (v1)
  (v0) edge[bend left] node[draw=none,below] {-1} (v3)
  (v1) edge node[draw=none,above] {-1} (v2)
  (v2) edge node[draw=none,below] {-1} (v3)
  (v3) edge node[draw=none,below] {1} (v4)
  (v4) edge node[draw=none,above] {1} (v1)
  (v0) edge[bend left] node[draw=none,above] {1} (v5)
  (v0) edge[bend right] node[draw=none,below] {-1} (v7)
  (v5) edge node[draw=none,above] {0} (v6)
  (v6) edge node[draw=none,below] {-1} (v7)
  (v7) edge node[draw=none,below] {0} (v8)
  (v8) edge node[draw=none,above] {1} (v5)
  ;

\end{tikzpicture}
\caption{A one-person game with 5 plays.}
\label{f1}
\end{figure}

\paragraph{Motivation.} 
The $0$-total reward is a classic payoff function \cite{Gilette57,LL69}. As we will show, the family of $1$-total reward games includes the family of 0-total reward games (see Corollary~\ref{c-k<k+1}). 
On the other hand, Thuijsman and Vrieze \cite{TV87,TV98} regard the $1$-total reward as a refinement of the $0$-total one, in the sense that it can be used to distinguish between optimal pairs of strategies which have the same $0$-total values (equal to $0$). In other words, the $1$-total reward, when used as a refinement of the $0$-total reward, has higher sensitivity, see e.g. \cite{P05}.

Furthermore, $1$-total reward games  have numerous applications, e.g., shortest path interdiction \cite{FH77,IW02,KGZ06, KBBEGRZ08} and scheduling with and/or precedence constraints \cite{MSS04}. 

Motivated by this, we introduce the $k$-total reward as a refinement of the $(k-1)$-total reward, and show a similar sensitivity behavior as in the case $k=1$. In fact, the larger $k$ is, the more the effective reward $\phi^{(k)}$ emphasizes the initial segment of the play. To understand this, it is instructive to look at the example given above.
One can imagine a scenario in which the 5 given sequences correspond to different possible plays of a BW-game (see Figure~\ref{f1}). Note that all the above sequences have equal values according to the $0$-total measure.  How to rank these plays? The $1$-total reward can distinguish between $\ba^1$ and $\ba^2$, for example, but not between $\ba^0$, $\ba^3$ and $\ba^4$. On the other hand, $2-$total reward can differentiate all the above $5$ sequences. 

\begin{remark}
The  $k$-total effective reward looks somewhat similar  
to the moment $M_k$ in the theory of probability. 
The first two cases  $k=0$ and $k=1$  
are easy to interpret and they play a very important role: 
the first two moments  $M_0$  and  $M_1$  are the expectation and variance, 
the $0$- and $1$-total payoffs are the mean and total effective payoffs, respectively. 
Yet, the higher moments have some important applications too.   
\end{remark}
\begin{remark}
The existence of optimal pure stationary strategies in $1$-total reward games can be derived from the general results in \cite{GW04}. The authors also left an open question at the end of the paper about effective reward functions $\phi$ that guarantee the existence of optimal pure stationary strategies for all BW-games, namely, whether such a $\phi$ can always be expressed as a non-decreasing function of a fairly mixing mapping of the reward sequence; see \cite{GW04} for the definition and more details. It is easy however to check that the $2$-total reward answers this question {\it in the negative}. 
\end{remark}

\paragraph{Main Results.}
First, we show that every $k$-total reward BW-game, when restricted to pure stationary strategies, has the same optimal strategy as the corresponding discounted mean payoff ($0$-total) game, if
the discount factor is close enough to $1$. Let us note however that this does not prove the existence of a saddle point among all (not necessarily pure or stationary) strategies. For the latter, one needs to show that in a $k$-total reward game, to a fixed pure stationary strategy of one player, there exists a best response of the other player that is also pure and stationary. In fact we show a stronger result, namely that there exists a pure stationary best response that {\it uniformly} dominates any other response; c.f. \cite{MN81} for the concept of uniformity.
This is our second result, which  with the previous claim implies now the existence of a saddle point in any $k$-total reward game. Moreover, this also shows that such a saddle point can be realized in uniformly optimal pure stationary strategies.  

\begin{remark}
Thuijsman and Vrieze \cite{TV98} gave a necessary and sufficient condition for a $1$-total general stochastic game to have a saddle point provided the players are restricted to their pure stationary strategies.  For the case of games with perfect information, this result was extended in \cite{BWR-1} to include history-dependent strategies. In particular, this gives another proof of the $1$-total case considered in the current paper. 
However, for $k>1$ no such condition is known.
\end{remark}

Next, we prove that the $k$-total reward BW-games can be embedded\footnote{that is, for every $k$-total reward game we can construct an equivalent $(k+1)$-total reward game, i.e.,  solving the latter provides a solution to the former.}
into the family of $(k+1)$-total reward BW-games, for each $k \in \ZZ_+$.
In particular,  mean payoff games can be embedded into $k$-total reward games for all  $k \in \ZZ_+$.
This containment and the example in \cite{Gurvich88} prove that
for each $k \in \ZZ_+$, there is a
non-zero sum $k$-total reward game without Nash equilibria.

%
%
%
%
%
%
%
%

\section{Iterated Total Rewards}\label{basic}\label{s-definitions-and-notations}

In this section we introduce a complete hierarchy of effective reward functions providing a natural generalization of mean and total payoffs, and their discounted counterparts.

Let us first introduce three operators acting on infinite sequences of reals. The \emph{limiting average} $A:\cS\rightarrow \RR$ and \emph{discounted limiting average} $A_{\gb}:\cS\rightarrow \RR$ operators map an infinite sequence into the set of reals, while the \emph{moment} $M:\cS\rightarrow \cS$ operator maps it into another infinite sequence.

More precisely, given an infinite sequence $\ba=(a_1,a_2,\ldots)$ and a real $0<\gb < 1$ we define
\begin{equation}\label{e-A}
A(\ba) ~=~ \liminf_{T\rightarrow \infty} \frac{1}{T} \sum_{j=1}^Ta_j
\end{equation}
and
\begin{equation}\label{e-Abeta}
A_{\gb}(\ba) ~=~ (1-\gb)\sum_{j=1}^{\infty} \gb^{j-1} a_j.
\end{equation}
For convenience, we also extend the definition of the operator $A$ for finite sequences $\ba$ to denote the average of the elements of $\ba$.

Finally, recall that we define $M(\ba)=\bb=(b_1, b_2,\ldots)\in\cS$ by setting
\begin{equation}\label{e-M}
b_i ~=~ \sum_{j=1}^i a_j ~~~~ \text{ for all } ~ i=1,2,\ldots
\end{equation}
For $k=0,1,\ldots,$ we call $M^k(\ba):=M(M^{k-1}(\ba))$ the \emph{$k$th moment sequence} of $\ba$, and define $M^0(\ba)=\ba$. For convenience we also extend $M$ for finite sequences. For $\ba\in\RR^n$, we use \raf{e-M} to define $\bb=M(\ba)\in\RR^n$.

We also introduce the following families of functions $\phi^{(k)}:\cS\rightarrow \overline\RR$ and $\phi^{(k)}_{\gb}:\cS\rightarrow \RR$ for $k=0,1,\ldots$, defined by
\begin{equation}\label{e-totalrewards}
\phi^{(k)}(\ba) ~=~ A\left(M^k(\ba)\right)  ~~~~~\text{ and} ~~~~~ \phi^{(k)}_{\gb}(\ba) ~=~ A_{\gb}\left(M^k(\ba)\right).
\end{equation}

Let us note that $\phi^{(0)}=\phi$ is the mean payoff function, $\phi^{(0)}_{\gb}=\phi_{\gb}$ is the discounted mean payoff, while $\phi^{(1)}=\psi$ is the total reward. Following this terminology, we call $\phi^{(k)}$ the \emph{$k$-total reward},
and $\phi^{(k)}_{\gb}$ the \emph{discounted $k$-total reward}. Thus the above hierarchy of effective reward functions provides a natural generalization of mean payoff and total reward.

\bigskip

We show first that the $M$ operator changes the discounted total rewards by a factor that depends only on $\beta$.

\begin{fact}\label{f-2b}
For all $\ba\in\cS$ and for all $0<\gb <1$ we have
\[
A_\gb(M(\ba)) ~=~ \frac{1}{1-\gb} A_\gb (\ba).
\]
\end{fact}

\proof
Using definition \eqref{e-Abeta} and \eqref{e-M} we can write
\[
\begin{array}{rl}
A_\gb(M(\ba)) &=\displaystyle (1-\gb)\sum_{i=1}^\infty \gb^{i-1}\left(\sum_{j=1}^i a_j\right)\\
&=\displaystyle (1-\gb)\sum_{j=1}^\infty a_j\left(\sum_{i=j}^\infty \gb^{i-1}\right)\\
&=\displaystyle (1-\gb)\sum_{j=1}^\infty \gb^{j-1}a_j\left(\sum_{i=j}^\infty \gb^{i-j}\right)\\
&=\displaystyle (1-\gb)\sum_{j=1}^\infty \gb^{j-1}a_j\left(\sum_{\ell =0}^\infty \gb^{\ell}\right)\\
&=\displaystyle (1-\gb)\sum_{j=1}^\infty \gb^{j-1}a_j\left(\frac{1}{1-\gb}\right)\\
&=\displaystyle \frac{1}{1-\gb} A_\gb (\ba).
\end{array}
\]
\qed

The above fact shows that $\phi_\gb^{(k)}(1-\gb)^k=\phi_\gb$, that is, all $k$-total reward BW-games have equivalent discounted versions. 


Given two sequences, $\bx\in\RR^p$ and $\by\in\RR^q$, let us
denote by $\ba=(\bx(\by))$ the infinite sequence obtained by listing first the elements of $\bx$ and
then repeating $\by$ cyclically, infinitely many times.
Let us call such an $\ba$ a \emph{lasso sequence}, and let us denote the set of lasso sequences that
can arise from a graph on $n$ vertices by
\[
\cS_n(R) ~=~ \left\{ \ba=(\bx(\by))\left| \begin{array}{cc}
p,q\in\ZZ_+, &p+q\leq n, ~q\ge 1\\*[3mm] \bx\in [-R,R]^p, &\by\in [-R,R]^q\end{array}\right.\right\},
\]
where $R$ is a given constant.
Note that a BW-game with $n$ positions in pure stationary strategies always produces a play $\bp$ such that
the corresponding rewards sequence $r(\bp)$ belongs to $\cS_n(R)$, if $R$ is an upper bound
on the absolute values of the local rewards.
We shall simply write $\cS_n$ when $R$ is not specified. In the sequel, when we write $\bx(\by)$ we assume that there are nonnegative integers $p$ and $q\ge 1$ such that $\bx\in\RR^p$, $\by\in\RR^q$, and $p+q\le n$. 

\bigskip

To be able to state and prove our main results about iterated total rewards, we need next to analyze the above operations on lasso sequences.

\begin{fact}\label{f-3}
For $\ba=(\bx(\by))\in \cS_n$, we have
\[
a_i=\left\{\begin{array}{ll}
x_i&\text{ if } i\leq p,\\*[3mm]
y_r&\text{ if } i=p+\ell q +r \text{ for some integers } \ell\geq 0, ~ 0< r\leq q.
\end{array}\right.
\]
\end{fact}

\begin{fact}\label{f-4}
For $\ba=(\bx(\by))\in \cS_n$, we have
\[
M(\ba)_i ~=~ \left\{\begin{array}{l@{\text{ if }}l}
\displaystyle\sum_{j=1}^i x_j& i\leq p\\*[3mm]
\displaystyle\sum_{j=1}^p x_j +\ell\sum_{j=1}^q y_j +\sum_{j=1}^r y_j & i=p+\ell q +r \text{ for integers } \ell\geq 0, ~0< r\leq q.
\end{array}
\right.
\]
\end{fact}

\begin{fact}\label{f-2a}
For $\ba=(\bx(\by))\in \cS_n$, we have 
$$
A(M(\bx(\by)))=\left\{\begin{array}{l@{\text{ if }}l}+\infty & A(\bx(\by))=A(\by)>0,\\-\infty & A(\bx(\by))=A(\by)<0.\end{array}\right.$$
\end{fact}

Note that  there is an example in \cite{BWR-1} showing that Fact~\ref{f-2a} does not necessarily  hold  for reward sequences corresponding to non-stationary strategies. Furthermore, $M(\bx(\by))$ is not a lasso sequence, in general. However, the above facts, obtained by simple counting arguments, imply the following claim, the second part of which can be obtained from the first part by induction on $k$:

\begin{fact}\label{c-main}
If $\bx(\by)\in\cS_n(R)$ such that $A(\by)=0$, then $M(\bx(\by))=\widetilde{\bx}(\widetilde{\by})\in\cS_n(nR)$, where
$\widetilde{\bx} = M(\bx)$ and $\widetilde{\by}= pA(\bx) + M(\by).$
Furthermore, if $\bx(\by)\in\cS_n(R)$ such that $A(\bx(\by))=A(M(\bx(\by)))=\cdots$
$=A(M^{k-1}(\bx(\by)))=0$, then $M^k(\bx(\by))\in\cS_n(n^kR)$.
\end{fact}
Recall that by adding a scalar to a vector we mean incrementing all components of the vector by the same scalar value.

The above properties allow us to generalize an inequality between the discounted and undiscounted payoffs shown by \cite{ZP96} for  the mean payoff case.

\begin{lemma}\label{l-GZP}
If $\bx(\by)\in\cS_n(R)$ such that $A(\bx(\by))=A(M(\bx(\by)))=\cdots$\linebreak $=A(M^{k-1}(\bx(\by)))=0$, then
\[
|\phi^{(k)}(\bx(\by))-\frac{1}{(1-\gb)^k}\phi_\gb(\bx(\by))| ~\leq~ 2(1-\gb) n^{k+1}R.
\]
\end{lemma}

\proof
It was shown in \cite{ZP96} that for a lasso sequence $(\bx(\by))\in \cS_n(R)$ we have 
$$
|A(\bx(\by))-A_\gb(\bx(\by))|\leq 2(1-\beta)n R.
$$
Applying this for an arbitrary lasso sequence $\widetilde{\bx}(\widetilde{\by})\in\cS_n(n^kR)$ we get
\[
|A(\widetilde{\bx}(\widetilde{\by}))-A_\gb(\widetilde{\bx}(\widetilde{\by}))|\leq 2(1-\gb)n(n^kR)=2(1-\gb)n^{k+1}R.
\]
By Fact \ref{c-main} we have $M^k(\bx(\by))\in\cS_n(n^kR)$, thus applying the above for $\widetilde{\bx}(\widetilde{\by})=M^k(\bx(\by))$ we get
\[
|A(M^k(\bx(\by)))-A_\gb(M^k(\bx(\by)))|\leq 2(1-\gb)n^{k+1}R.
\]
By Fact \ref{f-2b} we have $A_\gb(M^k(\ba))=\frac{1}{(1-\gb)^k}A_\gb(\ba)$, and thus the above implies our claim.
\qed

\section{Uniform Optimality within Stationary Strategies}\label{sec:3}

In this section we introduce formal definitions and notation for pure stationary strategies and prove that any $k$-total BW-game has a uniformly optimal saddle point when restricted to this family of strategies.  
Any such strategy corresponds to a lasso sequence. We recall Shapley's result that a discounted game 
has a uniformly optimal saddle point in pure stationary strategies, and then prove that $\phi^{(k)}$ ranks the lasso sequences in agreement with $\phi_\gb$ if $\gb$ is sufficiently close to $1$.
  
Let us consider a BW-game $(G,r,\pi)$.
As before, we denote by $R$ the largest absolute value of a local reward 
\[
R ~=~ \max_{e\in E} |r(e)|.
\]
For a subset $F\subseteq E$ of the arcs of the directed graph $G=(V,E)$ we denote by $d^+_F(v)$ the out-degree of vertex $v\in V$ in the subgraph $(V,F)$. A subset $F\subseteq E$ of the arcs is a pure stationary strategy of \Min\ (resp., \Max) if $d^+_F(v)=1$ for all $v\in B$ (resp., for all $v\in W$). Let us denote by $\mS_B$ and $\mS_W$ the sets of pure stationary strategies of \Min\ and \Max, respectively.

Given a pair of pure stationary strategies, $\mb\in\mS_B$ and $\mw\in\mS_W$, we have a unique walk $e_1,e_2,\ldots,e_p,e_{p+1},\ldots,e_{p+q},e_{p+1},e_{p+2},\ldots,e_{p+q},\ldots$ from every initial position $v_0\in V$, which consists of an initial path $e_1,e_2,\ldots,e_p$ followed by a cycle $e_{p+1},\ldots,e_{p+q}$, which we traverse infinitely many times. We denote by $\bx=\bx(v_0;\mb,\mw)$ and $\by=\by(v_0;\mb,\mw)$ the corresponding reward sequences $\bx=(r(e_j)\mid j=1,\ldots,p)$ and $\by=(r(e_{p+j})\mid j=1,\ldots,q)$, and by $c(v_0;\mb,\mw)$ the payoff value corresponding to effective reward $\pi$:
\[
c(v_0;\mb,\mw) ~=~ \pi (\bx(\by)).
\]
We say that a pair of strategies $(\mb^*,\mw^*)$, $\mb^*\in\mS_B$, $\mw^*\in\mS_W$ is a \emph{saddle point} for initial position $v_0\in V$ if
\begin{equation}\label{e-NE}
c(v_0;\mb^*,\mw) ~\leq~ c(v_0;\mb^*,\mw^*) ~\leq c(v_0;\mb,\mw^*)
\end{equation}
hold for all $\mb\in\mS_B$ and $\mw\in\mS_W$. We say that $(\mb^*,\mw^*)$ is a \emph{uniform saddle point} if \eqref{e-NE} holds for all initial positions $v_0\in V$.

The following result follows essentially from \cite{Shapley53}.

\begin{fact}\label{f-dmpUNE}
A BW-game with the discounted mean payoff function $\pi=\phi_{\gb}$ has a uniform saddle point for all $0<\gb<1$,
\end{fact}

Given a positive integer $\ell$ and a local reward function $r$, let us define $\epsilon(\ell,r)>0$ as the smallest positive number that can arise as the absolute value of an integer linear combination of the $r$ values with coefficients not larger than $n^{\ell+1}$. 
Note that by the above definition the following relations hold for all $\ell$, $r$, and lasso sequences $\bx(\by),\widetilde\bx(\widetilde\by)\in\cS_n$:
\begin{align}\label{e1-}
&\epsilon(\ell,r)\ge\epsilon(\ell+1,r),\\
\label{e2-}
&\phi^{(\ell)}(\bx(\by))\ne 0\implies |\phi^{(\ell)}(\bx(\by))|\ge\frac{\epsilon(\ell,r)}{n},\\
\label{e3-}
&\phi^{(\ell)}(\bx(\by))\ne \phi^{(\ell)}(\widetilde\bx(\widetilde\by))\implies |\phi^{(\ell)}(\bx(\by))-\phi^{(\ell)}(\widetilde\bx(\widetilde\by))|\ge\frac{\epsilon(\ell,r)}{n^2}.
\end{align}
Let us also note that if the local reward function $r$ is integral, then $\epsilon(\ell,r)\ge 1$ for all integers $\ell$.
 
We shall show next that an optimal pair of pure stationary strategies with respect to $\phi_\gb$ also forms a uniform saddle point with respect to the total reward $\pi=\phi^{(k)}$, if
$\gb$ is close enough to $1$.

\begin{theorem}\label{t-ktrUNE}
Consider $(G,r)$ as above, and choose a discount factor satisfying $0\leq (1-\gb) < \frac{\epsilon(k,r)}{4n^{k+3}R}$. Let us consider a uniform saddle point $(\mb^*,\mw^*)$ with respect to the discounted mean payoff $\phi_\gb$. Then, $(\mb^*,\mw^*)$ is also a uniform saddle point with respect to the $k$-total reward $\phi^{(k)}$.
\end{theorem}

\proof
We introduce two preorders $\prec_{\gb}$ and $\prec_k$ on the set of lasso sequences $\cS_n(R)$ as follows: for two sequences $(\bx(\by))$ and $(\widetilde\bx(\widetilde\by))$ we say that  $(\bx(\by))\prec_{\gb}(\widetilde\bx(\widetilde\by))$ (resp., $(\bx(\by))\prec_{k}(\widetilde\bx(\widetilde\by))$) if $\phi_\beta(\bx(\by)) \leq \phi_\beta(\widetilde\bx(\widetilde\by))$ (resp., $\phi^{(k)}(\bx(\by)) \leq \phi^{(k)}(\widetilde\bx(\widetilde\by))$).
We shall show that the preorder $\prec_{\gb}$ induced by $\phi_{\gb}$ on $\cS_n(R)$ is a refinement of $\prec_k$ induced by $\phi^{(k)}$. For this end let us first prove some partial claim about $\prec_{\gb}$. Introduce
\[
\begin{array}{r@{=~}l}
\cS(\ell,+) & \left\{\bx(\by)\in\cS_n(R)\mid \phi^{(j)}(\bx(\by))=0 ~\text{ for }~ j=0,1,\ldots,\ell-1, \text{ and } \phi^{(\ell)}(\bx(\by))>0\right\}\\*[3mm]
\cS(\ell,-) & \left\{\bx(\by)\in\cS_n(R)\mid \phi^{(j)}(\bx(\by))=0 ~\text{ for }~ j=0,1,\ldots,\ell-1, \text{ and } \phi^{(\ell)}(\bx(\by))<0\right\}
\end{array}
\]
for $\ell=0,1,\ldots,k-1$, and set
\[
\cS^* ~=~ \left\{\bx(\by)\in\cS_n(R)\mid \phi^{(j)}(\bx(\by))=0 ~\text{ for }~ j=0,1,\ldots, k-1\right\}.
\]
It is immediate to see by these definitions that these sets partition $\cS_n(R)$.
We claim that the following relations hold:
\begin{equation}\label{e-50}
\cS(0,-) \prec_{\gb} \cdots \cS(k-1,-)\prec_{\gb} \cS^*\prec_{\gb} \cS(k-1,+)\prec_{\gb} \cdots \prec_{\gb}\cS(0,+).
\end{equation}
To see this, let us fix an index $0\leq \ell <k$, and consider $\bx(\by)\in\cS(\ell,+)$ and $\widetilde{\bx}(\widetilde{\by})\in\cS(\ell+1,+)$ (or this could be $\cS^*$ if $\ell=k-1$).
Lemma \ref{l-GZP} implies that
\[
\left|\phi^{(\ell)}(\bx(\by))-\frac{\phi_{\gb}(\bx(\by))}{(1-\gb)^\ell}\right| ~\leq~ 2(1-\gb)n^{\ell+1}R<\frac{\epsilon(k,r)}{2n^{k+2-\ell}},
\]
where the last inequality follows from the choice of $\gb$.

By \raf{e1-} and \raf{e2-}, we have
\[
\phi^{(\ell)}(\bx(\by))\geq \frac{\epsilon(\ell,r)}{n}\ge\frac{\epsilon(k,r)}{n}.
\]
Thus, the above imply
\begin{equation}\label{e-51}
\frac{\phi_{\gb}(\bx(\by))}{(1-\gb)^\ell} > \frac{\epsilon(k,r)}{n}-\frac{\epsilon(k,r)}{2n^{k+2-\ell}}~>~ \frac{\epsilon(k,r)}{2n}.
\end{equation}
On the other hand we can apply Lemma \ref{l-GZP} for $\widetilde{\bx}(\widetilde{\by})$, too, yielding
\[
\left|\phi^{(\ell)}(\widetilde{\bx}(\widetilde{\by}))-\frac{\phi_{\gb}(\widetilde{\bx}(\widetilde{\by}))}{(1-\gb)^\ell}\right| ~\leq~ 2(1-\gb)n^{\ell+1}R<\frac{\epsilon(k,r)}{2n^{k+2-\ell}}.
\]
Since we have $\phi^{(\ell)}(\widetilde{\bx}(\widetilde{\by}))=0$,  we get
\begin{equation}\label{e-52}
\frac{\phi_{\gb}(\widetilde{\bx}(\widetilde{\by}))}{(1-\gb)^\ell}~<~\frac{\epsilon(k,r)}{2n^{k+2-\ell}} ~<~\frac{\epsilon(k,r)}{2n}.
\end{equation}
Inequalities \eqref{e-51} and \eqref{e-52} together imply that $\cS(\ell+1,+)\prec_{\gb}\cS(\ell,+)$ (or $\cS^*\prec_{\gb}\cS(k-1,+)$). Analogous arguments work for the negative side, too, and hence \eqref{e-50} follows.

Since $\phi^{(k)}$ has value $+\infty$ on $\cS(\ell,+)$ for $\ell=0,1,\ldots,k-1$ and value $-\infty$ on the sets $\cS(\ell,-)$ for $\ell=0,1,\ldots,k-1$ by Fact \ref{c-main}, the only thing left to prove that $\prec_{\gb}$ is a refinement of that of $\prec_k$ is to show that these two preorders do not conflict on the set $\cS^*$.

To this end let us consider two lasso sequences $\bx(\by),\widetilde{\bx}(\widetilde{\by})\in\cS^*$ such that $\phi^{(k)}$ has different values on these sequences. Since $\phi^{(k)}(\bx(\by))=A(M^k(\bx(\by)))$ and for $\bx(\by)\in\cS^*$ we have $M^k(\bx(\by))\in\cS_n(n^kR)$ by Fact~\ref{c-main}. We can conclude by \raf{e3-} that the difference between two non-equal $\phi^{(k)}$ values is always at least $\epsilon(k,r)/n^2$, say:
\[
\phi^{(k)}(\bx(\by)) - \phi^{(k)}(\widetilde{\bx}(\widetilde{\by})) ~\geq~ \frac{\epsilon(k,r)}{n^2}.
\]
As above, Lemma \ref{l-GZP} and our assumption on $\gb$ imply that
\[
\begin{array}{r@{~<~}l}
\left|\phi^{(k)}(\bx(\by))-\frac{\phi_{\gb}(\bx(\by))}{(1-\gb)^k}\right| & \frac{\epsilon(k,r)}{2n^2}, \text{ and}\\*[3mm]
\left|\phi^{(k)}(\widetilde{\bx}(\widetilde{\by}))-\frac{\phi_{\gb}(\widetilde{\bx}(\widetilde{\by}))}{(1-\gb)^k}\right| & \frac{\epsilon(k,r)}{2n^2}.
\end{array}
\]
The above three inequalities imply $\phi_{\gb}(\bx(\by)) > \phi_{\gb}(\widetilde{\bx}(\widetilde{\by}))$, proving our claim, and hence completing the proof of the theorem.
\qed

Note that the above result generalizes the result in \cite{ZP96} corresponding to the case $k=0$.
 
Let us remark that the proof of the above theorem in fact provides a more complete picture about these iterated total reward functions than the statement of the theorem alone. Namely, the effective reward functions $\phi^{(k)}$, $k=0,1,\ldots$ form a nested sequence, in the sense that the lasso sequences on which $\phi^{(k+1)}$ vanishes form a subset of those where $\phi^{(k)}$ vanishes, and $\phi^{(k+1)}$ has a finite value only on sequences on which $\phi^{(k)}$ vanishes.  Furthermore, \eqref{e-50} can be claimed for an arbitrary integer $k$, providing a complete hierarchy. 

\section{Best Response to Pure Stationary Strategies}
In this section, we  will show that, to a player's pure stationary strategy, the best response of the other player can be realized by a pure stationary strategy, even if that player is allowed to use any history dependent strategy. 

First we obtain several combinatorial results about finite sequences. Next, we will prove that any pure history dependent response to a pure stationary strategy can be replaced by a pure stationary strategy that is {\it uniformly} not worse, c.f. \cite{MN81}. Such uniformity allows us to prove that the best stationary response is at least as good as any history dependent  (not necessarily pure) strategy. 

We will use two operators on finite sequences: for a sequence $\ba=(a_1,a_2,\ldots, a_n)$ we have $S(\ba)=\sum_{i=1}^na_i$ and
$M(\ba)=(a_1,a_1+a_2,\ldots, S(\ba))$. Note that both $M:\RR^n\to\RR^n$ and $S:\RR^n\to\RR$ are linear operators.

\begin{lemma}\label{ln-0}
For sequences $\ba,\bb\in\RR^n$ and real $\gl\in\RR$ we have
\begin{subequations}
\begin{align}
S(\ba+\bb)&=S(\ba)+S(\bb)\label{en-2a}\\
S(\gl \ba)&=\gl S(\ba)\label{en-2b}\\
M(\ba+\bb)&=M(\ba)+M(\bb)\label{en-3a}\\
M(\gl \ba)&=\gl M(\ba)\label{en-3b}
\end{align}
\end{subequations}
\end{lemma}
\proof
It follows by the definitions of these operators.
\qed

To start our analysis, let us consider the all-one vector $\be=(1,1,\ldots,1)\in\RR^n$, and compute its iterated $M$ images and the corresponding sums.

\begin{lemma}\label{l-71}
For $\be=(1,1,\ldots,1)\in \RR^n$ and for every $k\in\ZZ_+$ we have
\begin{equation}\label{e-71a}
M^k(\be) ~=~ \left( \left. \binom{k-1+j}{k} ~\right|~ j=1,\ldots,n\right)
\end{equation}
and correspondingly
\begin{equation}\label{e-71b}
S(M^k(\be)) ~=~ \binom{n+k}{k+1}.
\end{equation}
\end{lemma}

\proof
The above expressions are clearly correct for $k=0$. We can prove them by induction on $k$ using the binomial identity
\begin{equation}\label{e-binom-1}
\sum_{j=0}^k \binom{a+j}{a} ~=~ \binom{a+k+1}{k}
\end{equation}
for all integers $a$ and $k$.
\qed

For a sequence $I$ of integers (indices), we denote by $\ba_I$ the sequence of the corresponding $\ba$ components. E.g., if $\ba=(a_1,a_2,a_3,\ldots)$ and $I=(1,2,3,2)$, then $\ba_I=(a_1,a_2,a_3,a_2)$. We denote by $[1,n]$ the sequence of integers from $1$ to $n$. If $\ba$ and $\bb$ are two finite sequences then we denote by $(\ba, \bb)$ their concatenation. For a sequence $I$ of indices we denote by $\be_I$ the sequence of $1$-s of length $|I|$ indexed by $i\in I$.

With this notation, the $k$-total value of an infinite sequence $\ba$ of local rewards can be rewritten as
\[
\phi^{(k)}(\ba) ~=~ \liminf_{T\rightarrow \infty} \frac{1}{T} S(M^k(\ba_{[1,T]})).
\]

Assume in the sequel that $X$, $Y$ and $Z$ are subsets of the indices, and $\ba$, $\bb$, and $\bc$ are finite sequences of appropriate lengths.

\begin{lemma}\label{ln-1}
\begin{subequations}
\begin{align}
M(\ba_{(X,Y)}) &=~ \left( M(\ba_X),~ M(\ba_Y)+S(\ba_X)\be_Y\right),\label{en-1}\\
M^k(\ba_{(X,Y)}) &=~ \left(M^k(\ba_X), M^k(\ba_Y) +\sum_{\ell=1}^k S(M^{k-\ell}(\ba_X))M^{\ell -1}(\be_Y)\right).\label{en-3}
\end{align}
\end{subequations}
\end{lemma}

\proof
Equality \eqref{en-1} follows by the definitions of $M$ and $S$. For \eqref{en-3} we use \eqref{en-1}, \eqref{en-3a}, \eqref{en-3b} and induction on $k$.
For $k=0$ we get $M^0(\ba_{(X,Y)})=(\ba_X,\ba_Y)$, and for $k=1$ we get
\[
M(\ba_{(X,Y)}) ~=~ M(\ba_X,\ba_Y)~=~ \left(M(\ba_X), M(\ba_Y) + S(\ba_X)\be_Y \right)
\]
by the definition of $M$, as in \eqref{en-1}.
Then, by induction on $k$ we get
\[
\begin{array}{rl}
M^{k+1}(\ba_{(X,Y)}) &= \displaystyle M\left(M^k(\ba_X), M^k(\ba_Y) +\sum_{\ell=1}^k S(M^{k-\ell}(\ba_X))M^{\ell -1}(\be_Y)\right)\\
&= \displaystyle \left( M^{k+1}(\ba_X), M^{k+1}(\ba_Y) + S(M^k(\ba_X))\be_Y + \sum_{\ell=1}^k S(M^{k-\ell}(\ba_X))M^{\ell}(\be_Y)\right)\\
&= \displaystyle \left( M^{k+1}(\ba_X), M^{k+1}(\ba_Y) + \sum_{\ell=1}^{k+1} S(M^{k+1-\ell}(\ba_X))M^{\ell-1}(\be_Y)\right).
\end{array}
\]
\qed

\begin{corollary}\label{cn-1}
\[
\begin{array}{rl}
S(M^k(\ba_{(X,Y)})) &=~ \displaystyle S(M^k(\ba_X))+ S(M^k(\ba_Y)) +\sum_{\ell=1}^k S(M^{k-\ell}(\ba_X))S(M^{\ell -1}(\be_Y))\\
&=~ \displaystyle S(M^k(\ba_X))+ S(M^k(\ba_Y)) +\sum_{\ell=1}^k S(M^{k-\ell}(\ba_X))\binom{|Y|+\ell-1}{\ell}.
\end{array}
\]
\end{corollary}
\proof
By Lemmas \ref{ln-0} and \ref{ln-1}, and by the equality
\begin{equation}\label{en-4}
S(M^{\ell -1}(\be_Y))~=~ \binom{|Y|+\ell-1}{\ell}.
\end{equation}
\qed

One of the operations we will use to prove our main claim of this section is the deletion of a middle interval from a sequence. To see the effect of such a step we need the following claims. 

\begin{corollary}\label{cn-2}
\[
\begin{array}{rrl}
M^k(\ba_{(X,Y,Z)}) &= \Biggl( & \displaystyle M^k(\ba_X), M^k(\ba_Y) + \sum_{\ell=1}^k S(M^{k-\ell}(\ba_X))M^{\ell-1}(\be_Y),\\
&& \hspace*{-1.5cm}\displaystyle M^k(\ba_Z) + \sum_{\ell=1}^k \left(S(M^{k-\ell}(\ba_X)) + S(M^{k-\ell}(\ba_Y)) \right) M^{\ell -1}(\be_Z)\\
&& \hspace*{-1cm} \displaystyle + \sum_{\ell =1}^k S(M^{k-\ell}(\ba_X))\sum_{m=1}^{\ell -1} \binom{|Y|+\ell -1-m}{\ell -m}M^{m-1}(\be_Z)~\Biggr).
\end{array}
\]
\end{corollary}
\proof
Apply Lemma \ref{ln-1} with $(Y,Z)$ instead of $Y$, and then again with $Y,Z$ instead of $X,Y$, and use
Lemma \ref{l-71}.
\qed

\begin{corollary}\label{cn-3}
\[
\begin{array}{c}
S(M^k(\ba_{(X,Y,Z)})) = S(M^k(\ba_X))+S(M^k(\ba_Y))+S(M^k(\ba_Z))\\*[5mm]
\displaystyle +\sum_{\ell =1}^k\left(S(M^{k-\ell}(\ba_X))\binom{|Y|+|Z|+\ell-1}{\ell} + S(M^{k-\ell}(\ba_Y))\binom{|Z|+\ell-1}{\ell}\right).
\end{array}
\]
\end{corollary}
\proof
Apply \eqref{en-2a} and \eqref{en-2b} of Lemma \ref{ln-0} and Corollary \ref{cn-2} above together with the equality \eqref{en-4} applied for both $\be_Y$ and $\be_Z$, and finally use the binomial identity
\begin{equation}\label{en-5}
\sum_{m=0}^{\ell}\binom{|Y|+\ell-1-m}{\ell-m}\binom{|Z|+m-1}{m} ~=~ \binom{|Y|+|Z|+\ell-1}{\ell}.
\end{equation}
\qed

\begin{corollary}\label{cn-4}
\[
S(M^k(\ba_{(X,Y,Z)})) = S(M^k(\ba_{(X,Z)})) + \sum_{\ell=0}^{k} \binom{|Z|+k-1-\ell}{k-\ell}\left(S(M^{\ell}(\ba_{(X,Y)}))-S(M^\ell(\ba_X))\right).
\]
\end{corollary}
\proof
Elementary calculations by Corollaries \ref{cn-1}, \ref{cn-3}, and by the binomial identity \eqref{en-5}.
\qed

\begin{corollary}\label{cn-5}
If $\ba=\bx(\by)$ is a lasso sequence, where $\bx=\ba_X$ and $\by=\ba_Y$, and
\begin{subequations}
\begin{align}
\left(S(M^{\ell}(\ba_{(X,Y)}))-S(M^\ell(\ba_X))\right)&=0 \text{ for } \ell=0,1,\ldots,k-1, \label{enz-1}
\end{align}
then we have
\begin{align}
\phi^{(k)}(\ba) &=~\frac{1}{|Y|}\left(S(M^{k}(\ba_{(X,Y)}))-S(M^k(\ba_X))\right).\label{enz-2}
\end{align}
Furthermore, if condition \eqref{enz-1} does not hold, and $0\leq m<k$ is the smallest index with $S(M^{m}(\ba_{(X,Y)}))-S(M^m(\ba_X))\neq 0$, then we have
\begin{align}
\phi^{(k)}(\ba) ~=~ \begin{cases} -\infty & \text{ if } ~~S(M^{m}(\ba_{(X,Y)}))-S(M^m(\ba_X))<0,\\
+\infty & \text{ if } ~~S(M^{m}(\ba_{(X,Y)}))-S(M^m(\ba_X))>0.
\end{cases}\label{enz-3}
\end{align}
\end{subequations}
\end{corollary}
\proof
For a positive integer $T$ let us define $\rho(T)=(T-|X|)\mod |Y|$ and $\ga(T)=\frac{T-|X|-r_T}{|Y|}$. Furthermore, let us denote by $Z_\rho$ the first $\rho$ elements of $Y$, for $0\leq \rho<|Y|$, and by $\ga Y = (Y,Y,\ldots,Y)$ the concatenation of $Y$ $\ga$ times. Then, for the lasso sequence $\ba=\ba_X(\ba_Y)$ we have
\begin{equation}\label{enz-3.5}
\phi^{(k)}(\ba) ~=~ \liminf_{T\rightarrow\infty}\frac{1}{T} S(M^k(\ba_{(X,\ga(T) Y, Z_{\rho(T)})}).
\end{equation}
Let us now apply Corollary \ref{cn-4} with $Z=((\lambda -1)Y,Z_{\rho(T)})$ for an arbitrary positive integer $\lambda$ to get
\begin{multline}\label{enz-4}
S(M^k(\ba_{(X,\lambda Y, Z_{\rho(T)})}) ~=~ S(M^k(\ba_{(X,(\lambda-1) Y,Z_{\rho(T)})}) \\*[5mm]
\hspace*{1cm} + \displaystyle\sum_{\ell=0}^{k} \binom{\rho(T)+(\lambda -1)|Y|+k-1-\ell}{k-\ell}\left(S(M^{\ell}(\ba_{(X,Y)}))-S(M^\ell(\ba_X))\right).
\end{multline}
Assume first that condition \eqref{enz-1} holds. Then the above equality simplifies to
\[
S(M^k(\ba_{(X,\lambda Y, Z_{\rho(T)})}) ~=~ S(M^k(\ba_{(X,(\lambda -1) Y, Z_{\rho(T)})}) + S(M^{k}(\ba_{(X,Y)}))-S(M^k(\ba_X)).
\]
Summing up the above equalities for $\lambda=1,\ldots,\ga(T)$, we get
\begin{multline}\label{enz-5}
S(M^k(\ba_{(X,\ga(T) Y, Z_{\rho(T)})}) ~=~ S(M^k(\ba_{(X, Z_{\rho(T)})}) \\+ \ga(T)\left(S(M^{k}(\ba_{(X,Y)}))-S(M^k(\ba_X))\right).
\end{multline}
Since $S(M^k(\ba_{(X, Z_{\rho(T)})})$ and $|X|+\rho(T)$ are both bounded independently of $T$, and since 
\[
\lim_{T\rightarrow \infty}\frac{\ga(T)}{T}=\lim_{T\rightarrow \infty}\frac{\ga(T)}{|X|+\ga(T)|Y|+\rho(T)}=\frac{1}{|Y|},
\]
we get from \eqref{enz-3.5} and \eqref{enz-5} that 
\[
\phi^{(k)}(\ba)=\frac{1}{|Y|}\left(S(M^{k}(\ba_{(X,Y)}))-S(M^k(\ba_X))\right)
\]
as claimed in \eqref{enz-2}.

\medskip

Let us assume next that condition \eqref{enz-1} does not hold, and let $m$ be the smallest index such that $S(M^m(\ba_{(X,Y)}))-S(M^m(\ba_X))\neq 0$. 
Consider equality \eqref{enz-4}, and note that the summation on the right hand side can be viewed as a polynomial $p(\lambda)$, which is of degree $k-m$ and in which the sign of the leading term is the same as the sign of $S(M^m(\ba_{(X,Y)}))-S(M^m(\ba_X))$. Then,
\[
\sum_{\lambda=1}^{\ga(T)} p(\lambda) = q(\ga(T))
\]
is a polynomial of $\ga(T)$ of degree $k-m+1$, and the sign of its leading term is the same as the sign of $S(M^m(\ba_{(X,Y)}))-S(M^m(\ba_X))$.
Thus, by summing up \eqref{enz-4} for $\lambda=1,\ldots,\ga(T)$ we get 
\begin{equation}\label{enz-6}
S(M^k(\ba_{(X,\ga(T) Y, Z_{\rho(T)})}) ~=~ S(M^k(\ba_{(X, Z_{\rho(T)})}) + q(\ga(T)).
\end{equation}
Since $k-m+1>1$ and since $S(M^k(\ba_{(X,Z_{\rho(T)})})$ and $|X|+\rho(T)$ are both bounded independently of $T$, we obtain \eqref{enz-3} from \eqref{enz-3.5} and \eqref{enz-6} by dividing by $T$ and taking limits.
\qed

Let us call in the sequel a lasso sequence {\it good} if \raf{enz-1} holds, and call it {\it bad} otherwise.

\bigskip

Given a BW-game $\Gamma$, we denote by $\mP_B(\Gamma)$ and $\mP_W(\Gamma)$ the sets of (not necessarily pure or stationary) strategies of \Min\ and \Max, respectively. Let us note that  the families of pure stationary strategies $\mS_B$ and $\mS_W$ are proper subsets of $\mP_B$ and $\mP_W$, respectively. Given a pair of strategies $(\mb,\mw)$ we denote by $\ba(\mb,\mw)$ the corresponding infinite sequence of expected local rewards. 

\medskip

Let us note that a pair of pure strategies defines an infinite walk in $G$ starting from the initial position $v_0$. In the next lemma we will analyze the structure of this walk and show that it can be decomposed in a certain way into lassos and paths. Let us introduce some notation first. 


Consider such an infinite walk $W=(v_0,v_1,\ldots)$ in $G$. This infinite walk will have some repeated vertices and arcs. As before, let us denote by $e_t=(v_{t-1},v_t)$ the edges of this walk, and by $a_t=r(e_t)$ the local rewards, for $t=1,2,\ldots$, and set $F=\{e_1,e_2,\ldots\}$ to denote the sequence of arcs in this infinite walk in $G$. 
Denote for integers $i<j$ by $[i,j]$ the set of integers $\{i,i+1,\ldots,j\}$, and set $J_0=\{1,2,\ldots\}$ to denote the set of positive integers. To an increasing subset $I=\{i_1,i_2,\ldots\}\subseteq J_0$ of the indices, $i_1<i_2<\cdots$, we associate the edge set $F(I)=\{e_i\mid i\in I\}$ and we say that $F(I)$ is a \emph{walk} in $G$ if the endpoint of $e_{i_s}$ is the beginning of $e_{i_{s+1}}$, that is if $v_{i_s}=v_{i_{s+1}-1}$ for all $s=1,2,\ldots$. Note that if $s\in I$, then $F(I\cap [1,s])$ is also a walk in $G$.

Let us consider an arbitrary infinite increasing sequence $I$ of integers such that $F(I)$ is a walk in $G$ and let $q=q(I)$ be the smallest integer $q\in I$ such that vertex $v_q$ is repeated in the walk $F(I\cap [1,q])$. Let us denote then by $p=p(I)$ the unique index $p<q$, $p\in I$ for which $v_p=v_q$. Let us introduce $C(I)=I\cap [p+1,q]$, set $P(I)=I\cap [1,p]$ and define $J(I)=I\setminus [p+1,q]$. Observe that $F(P(I),C(I))$ is a lasso sequence with $F(C(I))$ as its cycle, and that $F(J(I))$ is again a walk in $G$.

Next, starting with $I=J_0$ and $F=F(J_0)$, let us define $p_s=p(J_{s-1})$, $q_s=q(J_{s-1})$, $P_s=P(J_{s-1})$, $C_s=C(J_{s-1})$, $L_s=(P_s,C_s)$ and set $J_s=J(J_{s-1})$, recursively for $s=1,2,\ldots$. Note that by the above observation $F(J_k)\subseteq F$ is a walk in $G$, $F(C_s)$ is a cycle, and $F(L_s)$ is a lasso with $F(C_s)$ as its cycle, for every index $s$.

For a positive integer $T$ let us define $s(T)$ as the largest index $s$ for which $q_s\leq T$. Furthermore, for any $s\leq s(T)$, let us introduce $Q_s=[q_s+1,T]$. Note that $F(P_{s(T)},Q_{s(T)})$ is a simple path in $G$ (that is, a path with no repeated vertices).

\begin{lemma}\label{mfo55}
Let $W$ be an infinite walk as above and $\ba$ be the corresponding infinite sequence of local rewards. Then
for every time horizon $T$ we have the following equality:
\begin{align}\label{e13}
S(M^k(\ba_{[1,T]})) ~=~ &S(M^k(\ba_{(P_{s(T)},Q_{s(T)})}))+\nonumber\\
&\sum_{j=1}^{s(T)} \sum_{\ell=0}^k\binom{T-q_j-1+k-\ell}{k-\ell}\left(S(M^\ell(\ba_{(P_j,C_j)}))-S(M^\ell(\ba_{P_{j}}))\right).
\end{align} 
\end{lemma}
\proof
Let us observe the following simple relations:
\begin{align}
[1,T] &= (P_1,C_1,Q_1) \textrm{ and } \label{en-7-1}\\
(P_j,Q_j) &= (P_{j+1},C_{j+1},Q_{j+1}) \textrm{ for all } j<s(T).\label{en-7-2}
\end{align}
By repeated applications of Corollary \ref{cn-4} for $(X,Y,Z)=(P_j,C_j,Q_j)$ for $j=1,2,...,s(T)$ based on the partitions in \eqref{en-7-1}-\eqref{en-7-2}, we obtain \raf{e13}.
\qed

To arrive to our main result of this section, let us first show that to a pure stationary strategy $\mb^*$ of \Min\ there exists a pure stationary response of \Max\ that majorizes all pure (history dependent) responses uniformly in time. 
Since we have only finitely many pure stationary responses of \Max, there is a best one $\mw^*$ that maximizes $\phi^{(k)}(\ba(\mb^*,\mw^*))$.  Depending on whether $\phi^{(k)}(\ba(\mb^*,\mw^*))$ is finite or infinite we will consider 3 cases. 

We start with the case when this value is $+\infty$. Clearly, no other strategy can be better than this.

Next we consider the case when $\phi^{(k)}(\ba(\mb^*,\mw^*))$ is finite. For this case we need to introduce some notation. Let us denote by $\cB(\Gamma)$ the set of reward sequences corresponding to paths of length at most $n=|V|$ arcs in our graph, and let $\Theta\ge 0$ be the smallest constant satisfying the following inequalities: 
\begin{equation}\label{Theta}
\max_{\mb\in\cB(\Gamma)} S(M^k(b))\le\Theta+i\cdot\phi^{(k)}(\ba(\mb^*,\mw^*)),\text{ for }i=1,\ldots,n.
\end{equation}
Let us consider a lasso reward sequence $\bx(\by)$ and associate to it a polynomial $p_{\bx,\by}(\tau)$ as follows:
\[
p_{\bx,\by}(\tau)=\sum_{\ell=0}^{k-1}\binom{\tau+k-\ell}{k-\ell}\left(S(M^\ell(\bx,\by))-S(M^\ell(\bx))\right).
\]
Let $\cP$ be the set of all such polynomials with negative leading coefficient (that is the coefficient of the highest degree term) that correspond to lasso sequences of $G$ as above. Since the set $\cP$ is finite there exist  $\epsilon>0$ and integer $\hat\tau$ such that
\[
\forall \tau\ge\hat\tau\text{ and }\forall p\in\cP:~p(\tau)\le-\epsilon\tau.
\]  
Finally, let $\cL$ be the set of lasso reward sequences in $G$, and define
\[
K=\max_{\stackrel{\tau\ge 0}{p\in\cP}}\{p(\tau)\}+\max_{\bc_X(\bc_Y)\in\cL}\{S(M^k(\bc_X,\bc_Y))-|Y|\phi^k(\ba(\mb^*,\mw^*))\}.
\]

Note that $\Theta$, $\epsilon$, $\hat\tau$, and $K$ all depend only on the graph $G$, the local rewards $r$, $k$, and the strategy $\mb^*$ of \Min, but not on the strategies of \Max.  Note also that we will need the above definitions only when $\phi^{(k)}(\ba(\mb^*,\mw^*))$ is finite.

\begin{lemma}\label{lem:mfo}
Let $\mb^*$ be an arbitrary pure stationary strategy of \Min, $\mw^*$ be a best pure stationary response of \Max, and $\ba^*=\ba(\mb^*,\mw^*)$ be the local reward sequence corresponding to this pair. Assume that $\phi^{(k)}(\ba(\mb^*,\mw^*))$ is finite. Consider an
arbitrary pure (possibly history dependent) strategy $\mw$ of \Max\ and the corresponding reward sequence $\ba=\ba(\mb^*,\mw)$. Then we have the following inequality, for every $\delta>0$ and every $T>\frac{1}{\delta}\left(\frac{2K^2}{\epsilon}+K\cdot\hat\tau +\Theta\right)$:
\begin{equation}\label{en-999}
\frac{1}{T}S(M^{(k)}(\ba_{[1,T]})) ~\leq~ \phi^{(k)}(\ba^*)+\delta.
\end{equation}
\end{lemma}
   
\proof
Let us consider the infinite walk $W$ corresponding to the pair $(\mb^*,\mw)$ and apply Lemma~\ref{mfo55}. We are going to estimate from above the terms on the right hand side of \raf{e13} considering separately the path and the good and bad lassos. For notational simplicity in this proof we shall write $M(I)=M(\ba_I)$ where $\ba$ is the reward sequence corresponding to walk $W$ and $I$ is a sequence of indices.

Let us start with the path and observe by the definition of $\Theta$ that
\begin{equation}\label{Nu101}
S(M^k(P_{s(T)},Q_{s(T)})\le|P_{s(T)}\cup Q_{s(T)}|\phi^k(\mb^*,\mw^*)+\Theta.
\end{equation}
For each lasso  $F(P_j,C_j)$, it will be convenient to introduce the term
\[
H_j=\sum_{\ell=0}^{k}\binom{T-q_j-1+k-\ell}{k-\ell}\left(S(M^\ell(P_j,C_j))-S(M^\ell(P_j))\right). 
\]
Let us next consider a good lasso $F(P_j,C_j)$. Then by Corollary~\ref{cn-5} and the choice of $\mw^*$ we get 
\begin{align}\label{Nu102}
H_j
=S(M^k(P_j,C_j))-S(M^k(P_j))
=|C_j|\cdot\phi^{(k)}(\ba_{P_j}(\ba_{C_j}))\le|C_j|\cdot\phi^{(k)}(\ba^*).
\end{align}
Finally, we consider bad lassos. Let $j_\ga$ be the index of a bad lasso $F(P_{j_\ga},C_{j_\ga})$, for $\ga=1,\ldots,D(T)$, where $D(T)$ is the number of bad lassos appearing in $W$ within the time horizon $T$. Then we claim that the following inequality holds
\begin{align}\label{Nu103}
\sum_{\ga=1}^{D(T)}H_{j_\ga}&\le\sum_{\ga=1}^{D(T)}|C_{j_{\ga}}|\phi^{(k)}(\ba^*)+T\delta-\Theta.
\end{align}
To see this, observe that by the definition of $K$, for $\alpha>D(T)-\hat\tau$ we have 
\begin{align}\label{Nu104}
H_{j_\ga} - |C_{j_{\ga}}|\phi^{(k)}(\ba^*)\leq K.  
\end{align}
For indices $\ga\le D(T)-\hat\tau$, we have $T-q_{j_\ga}-1\ge\hat\tau$ since the $q_j$ indices are pairwise distinct, and thus by the definition of $K$ and $\epsilon$ we can write 
\begin{align}\label{Nu105}
H_{j_\ga}- |C_{j_{\ga}}|\phi^{(k)}(\ba^*)\leq K-\epsilon(T-q_{j_\ga}-1).  
\end{align}
Summing up for all bad lassos we get from \raf{Nu104} and \raf{Nu105} the following inequality
\begin{align}\label{Nu106}
\sum_{\ga=1}^{D(T)}\left(H_{j_\ga} - |C_{j_{\ga}}|\phi^{(k)}(\ba^*)\right)\le K\hat\tau+\sum_{\ga=1}^{D(T)-\hat\tau}(K-\epsilon(T-q_{j_\ga}-1)).
\end{align}

Let us note that since the $q_j$ indices are pairwise distinct we have the inequalities $T-q_{j_\ga}-1\ge D(T)-\ga\ge D(T)-\hat\tau-\ga$, for all $\ga=1,\ldots,D(T)-\hat\tau$. Consequently, $\sum_{\ga=1}^{D(T)-\hat\tau}(T-q_{j_\ga}-1)\ge \binom{D(T)-\hat\tau}{2}$. Thus from \raf{Nu106} we get 
\begin{align}
\sum_{\ga=1}^{D(T)}\left(H_{j_\ga} - |C_{j_{\ga}}|\phi^{(k)}(\ba^*)\right)&\le K\cdot D(T)-\epsilon\binom{D(T)-\hat\tau}{2}\nonumber\\
&\le\frac{2K^2}{\epsilon}+\hat\tau K \label{Nu107}\\ 
&\le T\delta-\Theta.\label{Nu108}
\end{align}
Inequality \raf{Nu107} is obvious when $D(T)\le \frac{2K}{\epsilon}+\hat\tau$, and otherwise it follows from the fact that 
\[
\sum_{\ga=1}^{D(T)-\hat\tau}(K-\epsilon(D(T)-\hat\tau-\ga))\le 0
\]
when $D(T)>\frac{2K}{\epsilon}+\hat\tau$. Inequality \raf{Nu108} follows by the bound on $T$. 

Summing up inequalities \raf{Nu101},\raf{Nu102}, and \raf{Nu103} for all good lassos yields the statement.
\qed

Let us finally consider the case when  $\phi^{(k)}(\ba(\mb^*,\mw^*))=-\infty$.
Let $\bar\Theta\ge 0$ and $\bar K$ be defined by the following equations: 
\begin{equation*}\label{Theta}
\bar\Theta=\max_{\mb\in\cB(\Gamma)} S(M^k(b)).
\end{equation*}
\[
\bar K=\max_{\stackrel{\tau\ge 0}{p\in\cP}}\{p(\tau)\}+\max_{\bx(\by)\in\cL}\{S(M^k(\bx,\by))-S(M^k(\bx))\}.
\]

Note again that $\bar\Theta$, $\epsilon$, $\hat\tau$, and $\bar K$ all depend only on the graph $G$, the local rewards $r$, and $k$, but not on the strategies of \Max. 

\begin{lemma}\label{lem:-infty}
Let $\mb^*$ be an arbitrary pure stationary strategy of \Min, $\mw^*$ be a best pure stationary response of \Max, and $\ba^*=\ba(\mb^*,\mw^*)$ be the local reward sequence corresponding to this pair. Assume that  $\phi^{(k)}(\ba(\mb^*,\mw^*))=-\infty$. Consider an
arbitrary pure (possibly history dependent) strategy $\mw$ of \Max\ and the corresponding reward sequence $\ba=\ba(\mb^*,\mw)$. Then for every $N>0$ and $T>\frac{2n^2}{\epsilon}\left(\bar\Theta+\bar K+N+\epsilon (\hat\tau+1)\right)$, we have the following inequality:
\begin{equation}\label{en-9990}
\frac{1}{T}S(M^{(k)}(\ba_{[1,T]})) ~\leq~ -N.
\end{equation}
\end{lemma}
\proof
We prove this claim similarly as in the previous lemma.
We consider again the infinite walk $W$ corresponding to the pair $(\mb^*,\mw)$ and apply Lemma~\ref{mfo55} to the reward sequence $\ba=\ba(\mb^*,\mw)$. For simplicity, we write again $M(X)$ instead of $M(\ba_X)$. We are going to estimate from above the terms on the right hand side of \raf{e13} considering separately the path and the lassos. Let us note that in this case all lassos are bad. Let us also observe by the definition of $\bar\Theta$ that
\begin{equation}\label{Nu201}
S(M^k(P_{s(T)},Q_{s(T)})\le\bar\Theta.
\end{equation}
For convenience we introduce again for each lasso  $F(P_j,C_j)$ the term
\[
H_j=\sum_{\ell=0}^{k}\binom{T-q_j-1+k-\ell}{k-\ell}\left(S(M^\ell(P_j,C_j))-S(M^\ell(P_j))\right). 
\]
For a lasso $F(P_j,C_j)$ with $j\le s(T)-\hat\tau$, we have
\begin{align}\label{Nu204}
H_{j}\leq \bar K-\epsilon (T-q_j-1), 
\end{align}
since $T-q_j-1\ge\hat\tau$. For a lasso $F(P_j,C_j)$  with $j>s(T)-\hat\tau$, we have
\begin{align}\label{Nu205}
H_{j}\leq \bar K.
\end{align}
Summing up inequalities \raf{Nu201},\raf{Nu204}, and \raf{Nu205} for all lassos yields the statement by elementary calculations.
\qed


The above analysis shows that there exists a pure strategy stationary $\mw^*$ of \Max\ (in response to \Min's fixed pure stationary strategy $\mb^*$) that uniformly dominates any other pure strategy $\mw$:
\begin{equation}\label{uniform}
\liminf_{T\to\infty}\frac{1}{T}S(M^{(k)}(\ba(\mb^*,\mw)_{[1,T]}))\le \limsup_{T\to\infty}\frac{1}{T}S(M^{(k)}(\ba(\mb^*,\mw)_{[1,T]}))\le\phi^{(k)}(\ba(\mb^*,\mw^*)).
\end{equation}

\begin{corollary}\label{cor:uniform}
Inequalities \raf{uniform} hold for any (not necessarily pure or stationary) strategy $\mw$ of \Max.
\end{corollary}
\proof
If $\phi^{(k)}(\ba(\mb^*,\mw^*))=+\infty$ then there is nothing to prove. Otherwise, observe that the reward sequence corresponding to any strategy $\mw$ is a convex combination of reward sequences corresponding to pure strategies. Therefore \raf{uniform} must hold for this combination as well.
\qed

\begin{remark}
A similar concept of value uniformity was considered in \cite{MN81}. 
\end{remark}
Now we are ready to state the main result of this section.

\begin{theorem}\label{t-best-response}
In a $k$-total reward BW-game, given a pure stationary strategy of a player, there exists a pure stationary best response of the opponent among all (not necessarily pure or stationary) strategies.
\end{theorem} 
\proof
The claim that the best response of \Max\ is pure and stationary follows from Corollary \ref{cor:uniform}. For the case when \Min\ responds to \Max, we can also see the claim by~\raf{uniform} after reversing the signs of all the rewards and interchanging the roles of the players.  
\qed

\begin{remark}
Let us observe that the above proof goes through even if each player has an effective reward function which is either $\liminf$ or $\limsup$  (they may differ for the two players). It also goes through if we exchange the order of $\liminf$ ($\limsup$) and the expectation operator $\EE[\cdot]$ 
\end{remark}

\section{Hierarchy of $k$-Total Rewards} 
In what follows we shall show that any $k$-total reward can be viewed as a special case of $(k+1)$-total rewards. In particular, mean payoff games are a special case of $1$-total reward games, and in general of $k$-total reward games. 

To state and prove our results we need to obtain a functional relation between $M(\bx(\by))$ and the vectors $\bx$ and $\by$. For this we will need a number of combinatorial results relate to the $S$ and $M$ operators.

\begin{lemma}\label{l-72}
For $\bz=(z_1,z_2,\ldots,z_n)\in \RR^n$ and for every $k\in\ZZ_+$ we have
\begin{equation}\label{e-72a}
M^k(\bz) ~=~ \left( \left. \sum_{j=1}^i\binom{k-1+i-j}{k-1}z_j ~\right|~ i=1,\ldots,n\right)
\end{equation}
and correspondingly
\begin{equation}\label{e-72b}
S(M^k(\bz)) ~=~ \sum_{j=1}^n\binom{k+n-j}{k}z_j.
\end{equation}
\end{lemma}

\proof
For $k=0$, the above formula with an extended definition of the binomial coefficients \cite{Gould12} shows that $M^0$ is the identity operator as assumed.
For $k=1$ the above expressions coincide with the definitions of the $M$ and $S$ operators. Thus, by induction on $k$ we can write
\[
\begin{array}{r@{=~}l}
M^{k+1}(\bz) ~=~ M(M^k(\bz)) & \displaystyle\left(\left. \sum_{\ell=1}^i \sum_{j=1}^\ell \binom{k-1+\ell -j}{k-1} z_j~\right|~ i=1,\ldots,n\right)\\*[3mm]
& \displaystyle\left(\left. \sum_{j=1}^i z_j\sum_{\ell =j}^i \binom{k-1+\ell -j}{k-1} ~\right|~ i=1,\ldots,n\right)\\*[3mm]
& \displaystyle\left(\left. \sum_{j=1}^i\binom{k+i -j}{k}z_j ~\right|~ i=1,\ldots,n\right)
\end{array}
\]
where the second equality follows by \eqref{e-72a}, and the last one by \eqref{e-binom-1}.
Finally \eqref{e-72b} follows by \eqref{e-72a} and \eqref{e-binom-1}.
\qed

Let us recall a few combinatorial identities from \cite{Gould12}, which we shall need in the sequel.

\begin{lemma}\label{l-binom1}
\[
\sum_{u=0}^N (-1)^u\binom{N}{u}\binom{X-u}{R} ~=~ \binom{X-N}{X-R}.
\]
\end{lemma}

\begin{lemma}\label{l-binom2}
\[
\sum_{u=0}^N \binom{X+u}{u}\binom{Y+N-u}{N-u} ~=~ \binom{X+Y+N+1}{N}.
\]
\end{lemma}

Now we are ready to provide an algebraic description of the $M$ operator over the set of lasso sequences.

\begin{lemma}\label{t-M-lasso}
Let us consider integers $p,q>0$, $k\geq 0$ and a lasso sequence $\bx(\by)\in\cS_n(R)$ with $\bx\in\ZZ^p$ and $\by\in\ZZ^q$,  satisfying
\begin{equation}\label{e-COND}
\phi^{(\ell)}(\bx(\by))=0 ~~~\text{ for }~~~ 0\leq \ell \leq k-1.
\end{equation}
Then, we have
\begin{equation}\label{e-t-M-lasso}\small
M^k(\bx(\by)) ~=~ M^k(\bx)\left( \sum_{\ell=1}^k S\left(M^{k-\ell}(\bx)\right)M^{\ell-1}(\be) ~+~ M^k(\by) \right)
\end{equation}
where $\be=(1,1,\ldots,1)\in\ZZ^q$, and
\begin{equation}\label{e-phi-k}\small
\begin{array}{c}
\displaystyle\phi^{(k)}\left(\bx(\by)\right)~=~ A\left( \sum_{\ell=1}^k S\left(M^{k-\ell}(\bx)\right)M^{\ell-1}(\be) ~+~ M^k(\by) \right)\\*[3mm]
\displaystyle~=~ \frac{1}{q} S\left( \sum_{\ell=1}^k S\left(M^{k-\ell}(\bx)\right)M^{\ell-1}(\be) ~+~ M^k(\by) \right)\\*[5mm]
\displaystyle~=~ \frac{1}{q}\sum_{j=1}^p x_j\left[\binom{q+p-j+k}{k}-\binom{p-j+k}{k}\right]~+~ \frac{1}{q}\sum_{i=1}^q\binom{k+q-i}{k}y_i.
\end{array}
\end{equation}
\end{lemma}

\proof
Let us note first that for $k=0$ the condition \eqref{e-COND} is empty, and the summation in \eqref{e-t-M-lasso} is an empty sum, yielding $M^0(\bx(\by))=\bx(\by)$. Furthermore, for $k=1$ we have by Fact \ref{c-main} that $M(\bx(\by))=\widetilde\bx(\widetilde\by)$, where $\widetilde\bx=M(\bx)$ and $\widetilde\by=S(\bx)\be + M(\by)$, in agreement with \eqref{e-t-M-lasso}, since condition \eqref{e-COND} is equivalent with saying $S(\by)=0$ by definition \eqref{e-totalrewards}.
Thus, \eqref{e-t-M-lasso} follows by induction on $k$ using the linearity of $M$ as in Lemma \ref{ln-0}.

Finally, \eqref{e-phi-k} follows from \eqref{e-t-M-lasso} after applying the $S$ operator to both sides of \eqref{e-t-M-lasso} yielding the second line of \eqref{e-phi-k}. Then, by using the linearity of $S$ by Lemma \ref{ln-0}
and applying Lemmas \ref{l-71} and \ref{l-72} we get
\[
\begin{array}{c}
\displaystyle \frac{1}{q}\sum_{\ell=1}^k\binom{q+\ell-1}{\ell}\sum_{j=1}^p\binom{k-\ell+p-j}{k-\ell}x_j ~+~ \frac{1}{q}\sum_{i=1}^q\binom{k+q-i}{k}y_i\\*[5mm]
\displaystyle ~=~ \frac{1}{q}\sum_{j=1}^p x_j\left[\sum_{\ell=1}^k\binom{q+\ell-1}{\ell}\binom{k-\ell+p-j}{k-\ell}\right] ~+~ \frac{1}{q}\sum_{i=1}^q\binom{k+q-i}{k}y_i.
\end{array}
\]
Finally, applying Lemma \ref{l-binom2} with $u=\ell$, $N=k$, $X=q-1$ and $Y=p-j$ and subtracting the $u=0$ term from both sides we get the last line of \eqref{e-phi-k}.
\qed

\begin{remark}
Formula \raf{e-phi-k} shows that the value $\phi^{(k)}(\bx(\by))$ is a linear combination of the components of $\bx$ and $\by$. Furthermore, it can be verified that these linear combinations for $k=0,1,\ldots,n-1$, are linearly independent. Consequently, if $\phi^n$ takes a finite value on a lasso sequence, then all local rewards on this sequence must be equal to 0. On the other hand, there are BW-games such that the $\phi^{(n-1)}$ value, from a certain starting position, is finite and different from zero.  
\end{remark}

We need an additional technical lemma.

\begin{lemma}\label{l-binom4}
Let $X,k\ge 0$ be integers. Then
\begin{equation}\label{identity}
\sum_{j=0}^{\lfloor k/2\rfloor}(-1)^j 2^{k-2j}\binom{k-j}{j}\binom{X+k-j}{k-j}=\binom{2X+k+1}{k}.
\end{equation}
\end{lemma}
\proof
Let $g(X,k)$ denote the summation on the left hand side. Thus, we want to show that $g(X,k)=\binom{2X+k+1}{k}.$
We apply induction on $X\ge 0$. For $X=0$, we have $g(0,k)=k+1$ by (2.4) in \cite{Gould12}, and hence \raf{identity} holds in this case. We assume now that it holds for $X$, and verify it for $X+1$. 

First we can show the following claim.
\begin{claim}\label{cl1}
\begin{equation}\label{e1}
g(X+1,k)=\frac{1}{X+1}\sum_{j=0}^{\lfloor k/2\rfloor}(X+k+1-2j)g(X,k-2j).
\end{equation}
\end{claim}
\proof
\begin{eqnarray*}
g(X+1,k)&=&\sum_{j=0}^{\lfloor k/2\rfloor}(-1)^j 2^{k-2j}\binom{k-j}{j}\binom{X+1+k-j}{k-j}\\
&=&\sum_{j=0}^{\lfloor k/2\rfloor}(-1)^j 2^{k-2j}\binom{k-j}{j}\binom{X+k-j}{k-j}\cdot\frac{X+k+1-j}{X+1}\\
&=&\frac{X+k+1}{X+1}g(X,k)+h(k),
\end{eqnarray*}
where 
\begin{eqnarray}\label{e2}
h(k)&:=&-\sum_{j=0}^{\lfloor k/2\rfloor}(-1)^j 2^{k-2j}\binom{k-j}{j}\binom{X+k-j}{k-j}\cdot\frac{j}{X+1}\nonumber\\
&=&\sum_{j=0}^{\lfloor (k-2)/2\rfloor}(-1)^j 2^{(k-2)-2j}\binom{(k-2)-j}{j}\binom{X+(k-2)-j}{(k-2)-j}\cdot\frac{X+k-1-j}{X+1}\nonumber\\
&=&\frac{X+k-1}{X+1}g(X,k-2)+h(k-2). 
\end{eqnarray}
The claim follows by iterative application of \raf{e2}.
\qed

By induction, we have $g(X,k-2j)=\binom{2X+k-2j+1}{k-2j}$. Thus, it remains to prove the following claim. 
\begin{claim}\label{cl2}
\begin{equation}\label{e3}
\frac{1}{X+1}\sum_{j=0}^{\lfloor k/2\rfloor}(X+k+1-2j)\binom{2X+k-2j+1}{k-2j}=\binom{2(X+1)+k+1}{k}.
\end{equation}
\end{claim}
\proof
We use induction on $k$. The base cases $k=0$ and $1$ are easily verified. Assume the statement holds for all integers less than $k$.
Denote by $f(k)$ the left hand side of \raf{e3}. Then

\begin{eqnarray*}
f(k)&=&\frac{1}{X+1}(X+k+1)\binom{2X+k+1}{k}+f(k-2)\\
&=&\frac{1}{X+1}(X+k+1)\binom{2X+k+1}{k}+\binom{2(X+1)+(k-2)+1}{k-2}\\
&=&\frac{(2X+k+1)!}{k!(2X+3)!}[2(2X+3)(X+k+1)+k(k-1)]\\
&=&\binom{2(X+1)+k+1}{k}.
\end{eqnarray*}
\qed

%
%
%

For the main claim of this section, we introduce a split operation on sequences: to a given sequence $\bx=(x_1,x_2,\ldots)$, we associate $\bx^{(1)}=(x_1,-x_1,x_2,-x_2,\ldots)$.

\begin{theorem}\label{t-phik<phik+1}
For a nonnegative integer $k$ and any lasso sequence $\bx(\by)$ we have
\begin{equation}\label{e-81}
\phi^{(k+1)}\left(\bx^{(1)}\left(\by^{(1)}\right)\right) ~=~ 2^{k-1} \phi^{(k)}\left(\bx(\by)\right).
\end{equation}
\end{theorem}

\proof
We prove this claim by induction on $k$. For $k=0$, let us note that 
$
M(x_1,-x_1,x_2,-x_2,\ldots)= (x_1,0,x_2,0,\ldots)
$
implying that 
$\phi^{(1)}(\bx^{(1)})=\frac{1}{2}\phi^{(0)}(\bx)$. 

Let us next assume that we have the equalities
\[
\phi^{(\ell+1)}\left(\bx^{(1)}\left(\by^{(1)}\right)\right) ~=~ 2^{\ell-1} \phi^{(\ell)}\left(\bx(\by)\right)
\]
for all $0\leq \ell < k$. Thus, in particular, the signs of $\phi^{(\ell+1)}\left(\bx^{(1)}\left(\by^{(1)}\right)\right)$ and $\phi^{(\ell)}\left(\bx(\by)\right)$ are the same for all $\ell<k$. Therefore, by Fact \ref{f-2a} and the definition of $\phi^{(k)}$, both sides of \eqref{e-81} are simultaneously equal to $\pm\infty$, whenever $\phi^{(k)}\left(\bx^{(1)}\left(\by^{(1)}\right)\right)\neq 0$. Hence it is enough to prove the claim for lasso sequences satisfying
\begin{equation}\label{e-82}
\phi^{(\ell+1)}\left(\bx^{(1)}\left(\by^{(1)}\right)\right)~=~\phi^{(\ell)}\left(\bx(\by)\right) ~=~0 ~~~\text{ for all }~~~ \ell =0,1,\ldots,k-1.
\end{equation}
Under these conditions Lemma \ref{t-M-lasso} can be applied and we get for the split sequence $\bx^{(1)}\left(\by^{(1)}\right)$ that $2q\phi^{(k+1)}\left(\bx^{(1)}\left(\by^{(1)}\right)\right)$ is equal to
\[
\begin{array}{l}
\displaystyle \sum_{j=1}^p x_j\sum_{r=0}^1(-1)^{r}\left[\binom{2(q+p-j)+1-r+k+1}{k+1}-\binom{2(p-j)+1-r+k+1}{k+1}\right] \\*[5mm]
\displaystyle \hspace*{2in}~+~\sum_{i=1}^q y_i\sum_{r=0}^1(-1)^{r}\binom{2(q-i)+1-r+k+1}{k+1}\\*[7mm]
=~\displaystyle \sum_{j=1}^p x_j\left[\binom{2(q+p-j)+k+1}{k}-\binom{2(p-j)+k+1}{k}\right]~+~\sum_{i=1}^q y_i\binom{2(q-i)+k+1}{k}.
\end{array}
\]
Let us also note that under conditions \eqref{e-82} we can apply Lemma \ref{t-M-lasso} to $\phi^{(\ell)}\left(\bx(\by)\right)$ and express it as in \eqref{e-phi-k} for all $\ell < k$. Let us finally note that using these expressions and using Lemma \ref{l-binom4} three times, with $X=q+p-j$, $X=p-j$ and $X=q-i$ we get
\begin{equation}\label{e5}
2\phi^{(k+1)}\left(\bx^{(1)}\left(\by^{(1)}\right)\right)~=~\sum_{j=0}^k\ga_j\phi^{(j)}(\bx(\by)),
\end{equation}
where by Lemma~\ref{l-binom4}
\[
\ga_j=\left\{
\begin{array}{ll}
0&\text{ if }j<\frac{k}{2},\\
(-1)^{k-j} 2^{2j-k}\binom{j}{k-j}&\text{ otherwise}.
\end{array}
\right.
\]
By condition \eqref{e-82}, equality~\raf{e5} further simplifies to
\[
2\phi^{(k+1)}\left(\bx^{(1)}\left(\by^{(1)}\right)\right)~=~\ga_k\phi^{(k)}\left(\bx(\by)\right) ~=~ 2^k\phi^{(k)}\left(\bx(\by)\right)
\]
from which the statement follows.
\qed

The above theorem allows us to view the family of $k$-total reward games as a subfamily of  $(k+1)$-total reward games.
\begin{corollary}\label{c-k<k+1}
Given a $k$-total reward BW-game, one can construct in linear time an equivalent $(k+1)$-total reward BW-game.
\end{corollary}

\proof
Let us consider an arbitrary BW-game $\Gamma=(G,r,\phi^{(k)})$ and let us define its split, denoted by $\widetilde{\Gamma}=(\widetilde{G}=(\widetilde{B}\cup\widetilde{W},\widetilde{E}),\widetilde{r},\phi^{(k+1)})$, as the game obtained from $\Gamma$ by subdividing each arc of local reward $r(u,v)$ by a vertex $w=w_{uv}$ and defining $\widetilde{r}(u,w)=r(u,v)$ and $\widetilde{r}(w,v)=-r(u,v)$. Clearly, there is a one-to-one correspondence between the strategies in $\Gamma$ and $\widetilde{\Gamma}$. Then, it is also clear that the expected reward sequence arising from a play in $\widetilde{\Gamma}$ is the  split sequence of the reward sequence arising from the corresponding play in $\Gamma$. Theorem \ref{t-phik<phik+1} implies that the games $(G,r,\phi^{(k)})$ and $(\widetilde{G},\widetilde{r},\phi^{(k+1)})$ are equivalent; more precisely, the effective values of the corresponding plays are equal up to a multiplicative factor of $2^{k-1}$.
\qed

\section{Discussion}
One can think of at least three directions for generalizing the above results as discussed below. 

\medskip

Due to an example of a non-zero sum mean payoff BW-game \cite{Gurvich88} that
does not have a Nash equilibrium in pure stationary strategies, we can conclude by Theorem~\ref{t-phik<phik+1}  that, for any  $k \in \ZZ_+$, there are non-zero sum $k$-total reward games that have no such Nash equilibria either.
Let us note that the construction in Corollary~\ref{c-k<k+1} creates games in which every directed cycle has zero length.
Interestingly, in this case we are not aware of any Nash equilibrium free example with $1$-total effective reward and 
without directed cycles of zero length \cite{BEGM12}.


\medskip

It seems also possible (and important) to generalize the above
results to general stochastic games with perfect information, or in other words,
to replace the BW-model considered in this paper
by the BWR-model, where R stands for random positions (origins of moves of chance); this model is equivalent to the classical Gillette's model (see \cite{BEGM13}). For $k=1$, it was shown in \cite{BWR-1} that there always exists a saddle point, which can be realized by pure stationary uniformly optimal strategies. The similar question for $k>1$ remains open; moreover it does not seem easy to formulate Shapley's equations in this case.

\medskip

For the case of non-perfect information, Thuijsman and Vrieze \cite{TV98} gave a necessary and sufficient condition for a $1$-total general stochastic game to have a saddle point provided the players are restricted to their pure stationary strategies.  Extending these results to $k>1$ seems also hard.

\medskip


Finally, we discuss the complexity issues related to $k$-total reward BW-games. 
It is a long-standing open question whether there is a polynomial time algorithm for the case $k=0$, although it can be solved in pseudo-polynomial time \cite{Pis99,ZP96}
and randomized subexponential time \cite{BV05,Halman07,Vorobyov08}.
Let us also remark that the problem of deciding if the value of a mean payoff BW-game is below (or above) a given threshold belongs to both NP and co-NP
(\cite{GKK88, KL93, ZP96}).  Our reduction in Theorem~\ref{t-phik<phik+1} shows that solving $k$-total reward games is at least as hard as solving mean payoff games. In particular, every $0$-total game can be viewed as a special $1$-total game. Yet, a polynomial reduction in the other direction is not known. On the other hand, Theorem~\ref{t-ktrUNE} shows that the $k$-total BW-game with integral local rewards can be solved in pseudo-polynomial time\footnote{that is, the running time is bounded by a polynomial in $n$ and $R$.}. Furthermore, the 1-player case (that is, Markov decision processes) with $k\le 1$ can be solved in polynomial time via linear programming \cite{MO70,BWR-1}. The similar algorithmic question for $k>1$ is open.


\section*{Acknowledgments} We thank the two anonymous reviewers for the careful reading and many helpful remarks. Part of this research was done at the Mathematisches Forschungsinstitut Oberwolfach during a stay
within the Research in Pairs Program from July 26 to August 15, 2015. This research was partially supported by the Scientific Grant-in-Aid from Ministry of Education, Science,
Sports and Culture of Japan. The first author also thanks the
National Science Foundation (Grant IIS-1161476).  



\end{document}